# Nuclear Resonances, Scattering and Reactions from First Principles: Progress and Prospects

Sofia Quaglioni[1] and Petr Navrátil[2]
[1]Lawrence Livermore National Laboratory, P.O. Box 808, L-414, Livermore, CA 94551, USA
[2]TRIUMF, 4004 Wesbrook Mall, Vancouver, BC V6T 2A3, Canada

## Introduction

The study of rare isotopes at radioactive beam facilities has opened new frontiers in nuclear physics. To enable the discovery of model deficiencies and missing physics it is essential that the new insights from these experiments be confronted with predictive theoretical frameworks capable of describing the interplay of many-body correlations and continuum dynamics characteristic of exotic nuclei, as well as the nuclear reactions used to produce and study them. The predictive power garnered, in part, through this process is paramount to the use of atomic nuclei as a doorway to explore some of the most fundamental laws of the Universe through precision experiments. A primary example is neutrinoless double-beta ($0\nu\beta\beta$) decay. If observed, this exceedingly rare decay would provide definite evidence that neutrinos are their own antiparticles (i.e., Majorana particles), and a means to determining the neutrinos' masses. Essential for this latter goal are accurate predictions of $0\nu\beta\beta$ nuclear transitions, which can only be determined theoretically. A predictive theory of nuclear structural and reaction properties is also desirable to aid in precisely determining thermonuclear reaction rates that play an important role in fusion-energy experiments, the predictions of stellar-evolution models, and simulations of nucleosynthetic processes. Very difficult or even impossible to measure at the relevant energies of tens to hundreds of keV due to the hindering effect of Coulomb repulsion, thermonuclear fusion rates are almost always estimated by extrapolation from higher-energy experimental data. Without the help of a predictive theory, such extrapolations can be a major source of uncertainty. In addition to reducing the uncertainty in reaction rates at very low temperatures, a first-principles theory can also help resolving open issues such as the effect of polarization induced by the strong magnetic fields in plasma environments (such as at the international project ITER in France and at the National Ignition Facility in the USA) on the deuterium-tritium and deuterium-deuterium thermonuclear reaction rates.

## What is understood under *ab initio* nuclear theory?

At the low energies relevant for most studies of nuclear structure and nuclear reactions, currently the best path to achieving a predictive theory of nuclear properties combines effective field theories of quantum chromodynamics—that organize the nuclear force into systematically improvable expansions—with *ab initio* methods—that solve the quantum many-nucleon problem with controlled approximations.

In the last few years, the emergence of powerful ab initio approaches to nuclear structure has dramatically accelerated this journey, allowing for the description of nuclei as heavy as $^{100}$Sn.[1] Fewer ab initio efforts have been focused on attaining a unified description of bound-state and continuum properties and developing techniques applicable to nuclear reactions. Nevertheless, since the first ab initio calculation of neutron scattering on $^4$He in 2007[2], progress has been quite remarkable also in this direction. Large scale computations combined with new and sophisticated theoretical approaches have enabled high-fidelity predictions for nucleon- and deuterium-induced scattering and reactions on light targets and some medium-mass nuclei, and the *ab initio* description of $^3$H-$^4$He and $^3$He-$^4$He scattering and radiative capture processes (see Ref. [3] and references therein) as well as $^4$He-$^4$He scattering[4]. More recently, an avenue for arriving at the *ab initio* description of scattering and reactions in medium-mass nuclei has been opened by the development of optical nucleon-nucleus potentials[5].

## The *ab initio* no-core shell model with continuum

To arrive at an *ab initio* description of low-energy nuclear reactions, over the past 13 years we have been developing the no-core shell model with continuum (NCSMC), [6] a unified framework for the treatment of both bound and unbound states in light nuclei. With chiral two- (NN) and three-nucleon (3N) interactions as the only input, we are able to predict structure and dynamics of light nuclei and, by comparing to available experimental data, test the quality of chiral nuclear forces.

Describing a reaction–for example the scattering of $^3$He with $^4$He—requires addressing both the correlated short-range behavior occurring when the reactants are close together, forming a composite nucleus ($^7$Be in our example), and the clustered long-range behavior occurring when the reactants ($^3$He and $^4$He in our example) are far apart. The NCSMC accomplishes this by adopting a generalized cluster expansion for the wave function of the reacting system, which in the $^7$Be example is given by

$$\left|\Psi^{J^\pi}_{^7\mathrm{Be}}\right\rangle = \sum_\lambda c_\lambda^{J^\pi} \left|\,^7\mathrm{Be}, \lambda\, J^\pi\right\rangle + \sum_\nu \int dr\, r^2\, \gamma_\nu^{J^\pi}(r)\, \hat{\mathcal{A}}_\nu \left|\Phi^{J^\pi}_{\nu,r}\right\rangle. \tag{1}$$

In the first term, consisting of a expansion over (square-integrable) eigenstates of the composite system ($^7$Be) obtained within the no-core shell model (NCSM)[7] and indexed by $\lambda$, all $A$ nucleons are treated on the same footing. In the second term, corresponding to a resonating-group method[8] expansion over (continuous) antisymmetrized channels the wave function is factorized into products of cluster components ($^4$He and $^3$He) and their relative motion, with proper bound-state or scattering boundary conditions. Here, $\vec{r}_{4,3}$ is the separation between the centers-of-mass

$$\left|\Phi^{J^\pi}_{\nu,r}\right\rangle = \left[\left(\left|\,^4\mathrm{He}\right\rangle \left|\,^3\mathrm{He}\right\rangle\right)^{(s)} Y_\ell\left(\hat{r}_{4,3}\right)\right]^{J^\pi} \frac{\delta(r - r_{4,3})}{r r_{4,3}} \tag{2}$$

of $^4$He and $^3$He and ν is a collective index for the relevant quantum numbers. The discreet expansion coefficients $c_\lambda^{J^\pi}$ and the continuous relative-motion amplitudes $\gamma_\nu^{J^\pi}(r)$ are obtained as a solution of the generalized eigenvalue problem derived by representing the Schrödinger equation in the model space of expansion (1). The cluster eigenstates (e.g., $^4$He and $^3$He) are obtained within the NCSM with the same Hamiltonian used to describe the whole system. In general, the sum over $\nu$ includes also excited states of clusters, as well as different cluster partitions.

## Scattering and gamma-capture reactions

Nucleon elastic scattering on $^4$He is the most straightforward process to calculate within the NCSMC. The $^4$He nucleus, also known as $\alpha$ particle, is tightly bound with the lowest excited state at 20.2 MeV. Consequently, it is typically sufficient to consider only the $^4$He ground state in the generalized cluster expansion (1). Recently, $n$-$\alpha$ scattering became also treatable within the Faddeev-Yakubovsky formalism.[9] Benchmark calculations using the same chiral NN interaction as input demonstrated agreement between the computed phase shifts, providing validation for these two vastly different methods.[9] Having the capability to also include the chiral three-nucleon (3N) interaction in the calculations, we were able to reproduce the experimental phase shifts in the region of the $3/2^-$ resonance [located at about 0.8 (1.5) MeV above the $^4$He+n ($^4$He+p) threshold], corresponding to the ground state of $^5$He ($^5$Li). That allowed us to compute the $^4$He(p,p)$^4$He and $^1$H(α,p)$^4$He proton elastic scattering and recoil reactions, respectively, at energies of a few MeV per nucleon. These processes are the leading means for determining the concentrations and depth profiles of helium and hydrogen, respectively, at the surface of materials or in thin films. As seen in Fig. 1, our calculated cross sections are in a good agreement with experimental data.[10]

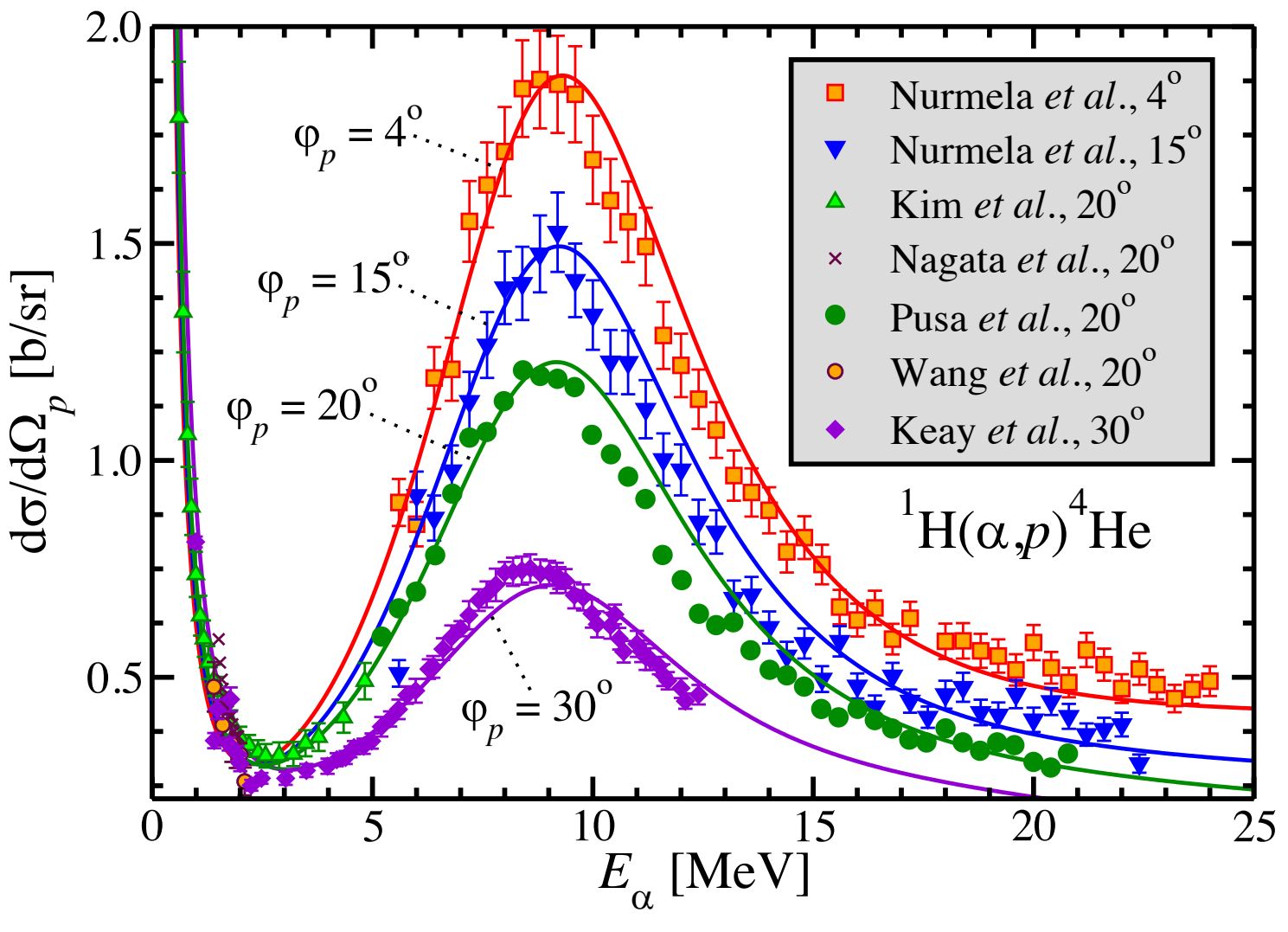

*Figure 1 Computed (lines) $^1$H(α,p)$^4$He angular differential cross section at proton recoil angles $\varphi_p$ = 4°, 15°, 20°, and 30° as a function of the incident $^4$He energy compared with data (symbols). Figure from Ref. [10].*

Nuclear scattering and reactions involving deuterium ($d$), the deformation and breakup of which cannot be neglected even at low energies due to its weak binding, pose an additional challenge. Deuteron-$\alpha$ elastic scattering is the simplest of such processes. We successfully applied the NCSMC to the $d$-$^4$He system and gained insight on the structure of $^6$Li resonances as well as of its bound ground state.[11] The deuteron deformation and breakup was taken into account by including proton-neutron excited pseudo states, i.e., a form of approximate discretization of the proton-neutron continuum. These calculations, carried out with chiral NN+3N forces provide a realistic description of $^{6}$Li and highlight the sensitivity of the system to the 3N interaction. Omitting the chiral 3N interaction results in significant overestimation of the excitation energy of the $3^{+}$ state, the lowest excitation of $^{6}$Li, as well as an underprediction of the splitting among the $3^{+}$-$2^{+}$-$1^{+}$ resonance triplet, dominated by d-$^{4}$He *D*-wave of relative motion coupled with the deuteron's *S*=1 spin. This demonstrates that the 3N interaction contributes substantially to the nuclear spin-orbit force strength. The calculated elastic differential cross section off the $3^{+}$ resonance region is described very well as seen in Fig. 2. The $^{6}$Li *S*- and *D*- wave asymptotic normalization constants, i.e., the normalizations of the wave function tail, are also close to the values inferred from analyses of experimental data. The S-wave asymptotic normalization constant determines the $^{2}$H(α,γ)$^{6}$Li radiative capture cross section, responsible for the Big-Bang nucleosynthesis of $^{6}$Li.

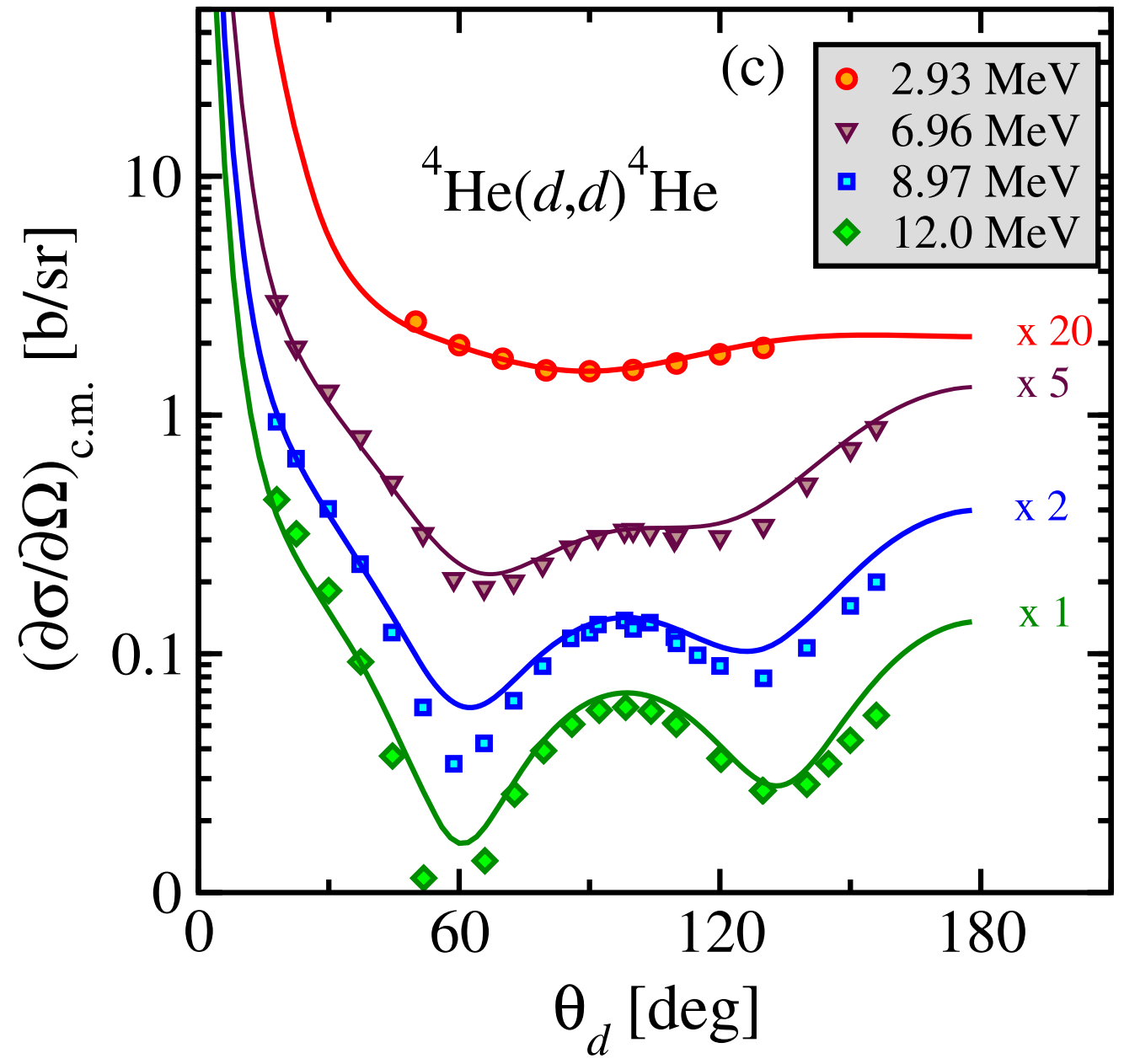

*Figure 2 Calculated (lines) and measured (symbols) center-of-mass angular distributions at $E_d$=2.93; 6.96; 8.97, and 12 MeV are scaled by a factor of 20, 5, 2, and 1, respectively. For details see Ref. [11].*

A well-known reaction important for both Big-Bang nucleosynthesis and the Solar proton-proton chain is the $^{3}$He(α,γ)$^{7}$Be radiative capture. Initial NCSMC calculations of $^{3}$He-$^{4}$He and $^{3}$H-$^{3}$He scattering[12] were carried out starting from a two-nucleon Hamiltonian. The properties of the low-lying resonances as well as those of the two bound states of $^{7}$Be and $^{7}$Li were reproduced rather well. With the obtained scattering and bound state wave functions, we also computed the astrophysical S-factor for the $^{3}$He(α,γ)$^{7}$Be solar fusion cross section (Fig. 3) as well as that of its mirror reaction $^{3}$H(α,γ)$^{7}$Li.[12] At very low energies, the $^{3}$He(α,γ)$^{7}$Be S-factor is in a good agreement with the measurements taken at the underground LUNA facility (Co07). However, its overall shape does not match some of the recent data, likely owing to an overestimation of the non-resonant $S$-wave phase shifts. A more quantitative prediction will require the inclusion of chiral 3N forces. One interesting observation is the that no microscopic theoretical approach is currently able to reproduce simultaneously the experimental normalizations of both the $^3$He(α,γ)$^7$Be and $^3$H(α,γ)$^7$Li S-factors. Our calculations overpredict the latter.[12]

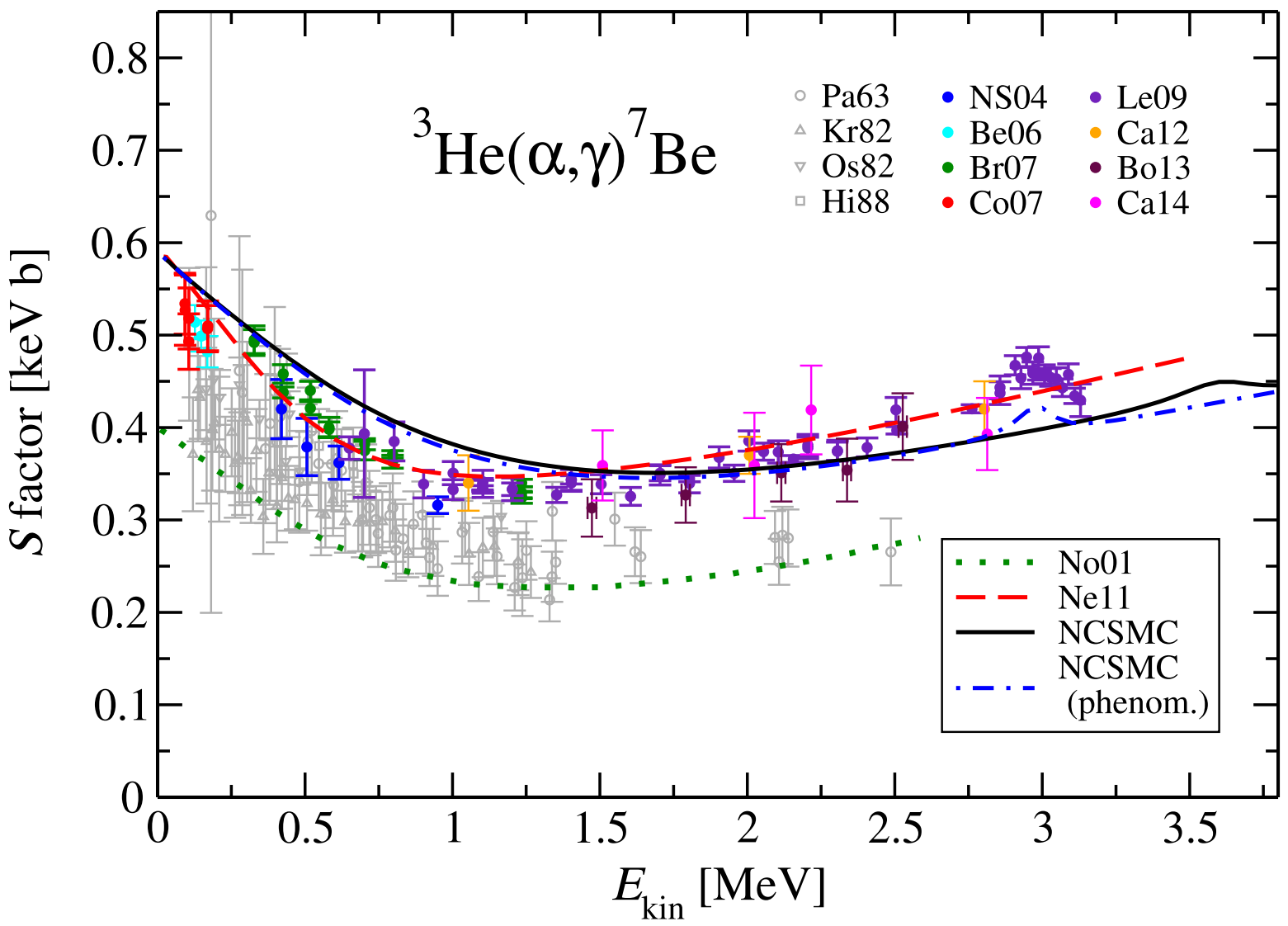

*Figure 3 Astrophysical S factor for the $^3$He(α,γ)$^7$Be radiative-capture processes obtained from the NCSMC approach compared with other theoretical approaches and with experiments. For details see Ref.[12].*

## Structure of weakly bound exotic nuclei

NCSMC calculations have also helped shedding light on the properties of halo nuclei, exotic weakly bound systems with one or two nucleons (typically neutrons) well-separated from the rest of the nucleus. One of the best examples is $^{11}$Be, famous for the "parity-inversion" of its ground and first excited states. With 7 neutrons, the standard shell model expects a $1/2^-$ ground state with the last neutron occupying the $0p_{1/2}$ level. However, experimentally $^{11}$Be has a $1/2^+$ ground state bound by only about 500 keV with respect to the $^{10}$Be+$n$ threshold. Understandably, the dynamics of the $^{10}$Be+$n$ system needs to be properly taken into account to realistically describe $^{11}$Be, something the NCSMC is perfectly suited for. Still, the details of the nuclear interaction continue to play a critical role and only a chiral force that describes well the nuclear density is capable to reproduce well the $^{11}$Be properties.[13] In Fig. 4, we show the overlap of the calculated $^{11}$Be $1/2^+$ ground-state wave function with $^{10}$Be+$n$ as a function of $^{10}$Be and neutron separation. The halo $S$-wave component (solid and dashed black lines) extends beyond 20 fm, i.e., very far beyond the range of the nuclear interaction. Also shown (dotted black line) is the $S$-wave overlap obtained from describing $^{11}$Be within the NCSM alone, i.e., without $^{10}$Be+n cluster component. While the spectroscopic factors obtained by integrating the square of the two overlap functions are about the same, the NCSM result is close to zero already at about 8 fm.

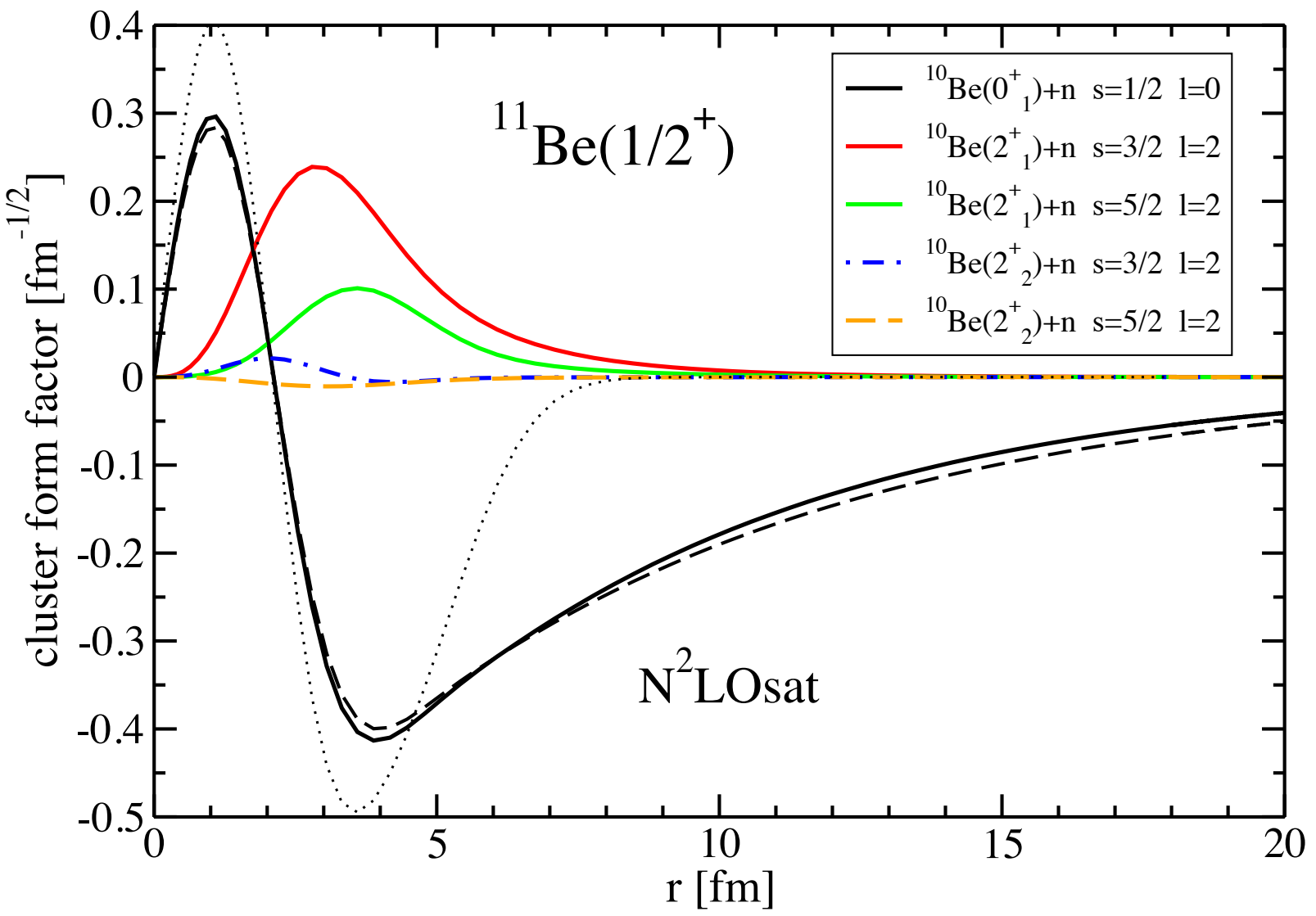

*Figure 4 Overlap of $^{11}$Be ground-state wavefunction with $^{10}$B+n as a function of $^{10}$Be and neutron separation. The black dashed line corresponds to the original NCSMC calculation for the S-wave, while the black full line is obtained after a phenomenological correction to fit the experimental threshold. The black dotted line shows the NCSM S-wave result that serves as one of the inputs into the NCSMC. For further details see Ref. [13].*

With the availability of the first re-accelerated $^{10}$C beam at TRIUMF, we teamed up with the TRIUMF IRIS collaboration and investigated the $^{10}$C(p,p)$^{10}$C elastic scattering and the structure of the unbound $^{11}$N nucleus.[14] As the collision energy was about 4 MeV in center of mass, the cross section was sensitive to the $5/2^+$ and $3/2^-$ resonances in $^{11}$N, analogs of well-known lowest-lying resonances in $^{11}$Be. Perhaps not surprisingly, the chiral NN+3N force that described successfully the parity inversion of the $^{11}$Be ground state also provided the best description of the measured differential cross section, although the overall normalization was overestimated as seen in Fig. 5. Interestingly this investigation demonstrated how an observable that is rather straightforward to measure, such as the elastic scattering cross section, can be a sensitive probe of nuclear force models.

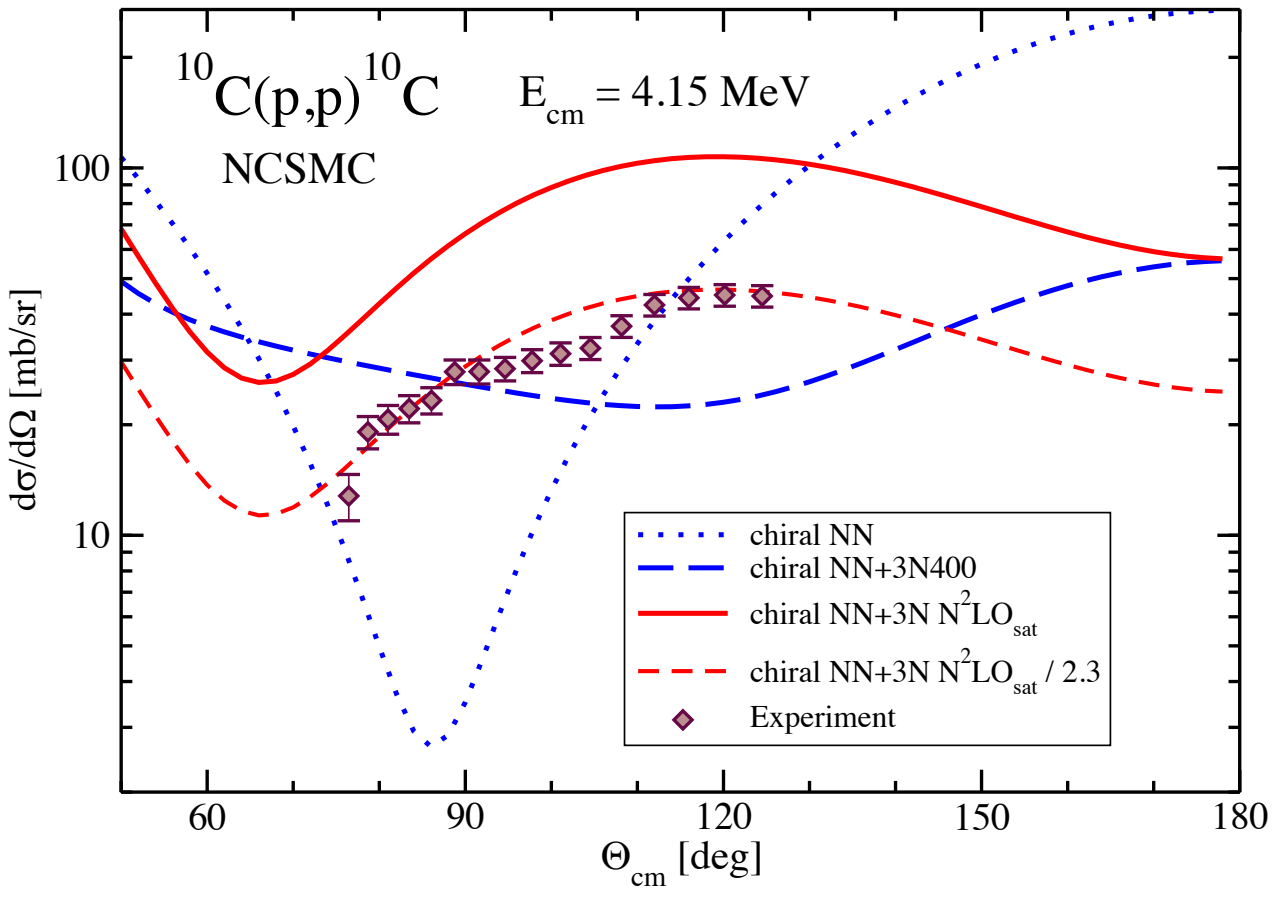

*Figure 5 Measured differential cross section of $^{10}$C(p,p)$^{10}$C(gs) at $E_{cm}$ = 4.15 MeV. The curves are ab initio theory calculations. For details see Ref.[14].*

More complex halo nuclei are Borromean systems with two loosely bound neutrons. The simplest example is $^6$He, which presents a $^4$He-n-n three-body cluster structure with none of the two-body subsystems being bound. We were able to generalize the NCSMC formalism to include three-body cluster dynamics.[15] The formalism and the computational effort become significantly more challenging. Compared to binary processes, the number of channels ($\nu$) increases dramatically. Further, the wave function has to be described up to very large hyperradial distances (on the order of 100 fm) to reach the asymptotic region. Because of the increased complexity, so far three-cluster NCSMC calculations do not include 3N forces. In spite of this, our calculations reproduce rather well the properties of the bound ground state of $^6$He (including the characteristic di-neutron and cigar configurations of the probability distribution, shown in Fig. 6, the charge radius and the two-neutron separation energy) as well as its low-lying resonances.[15]

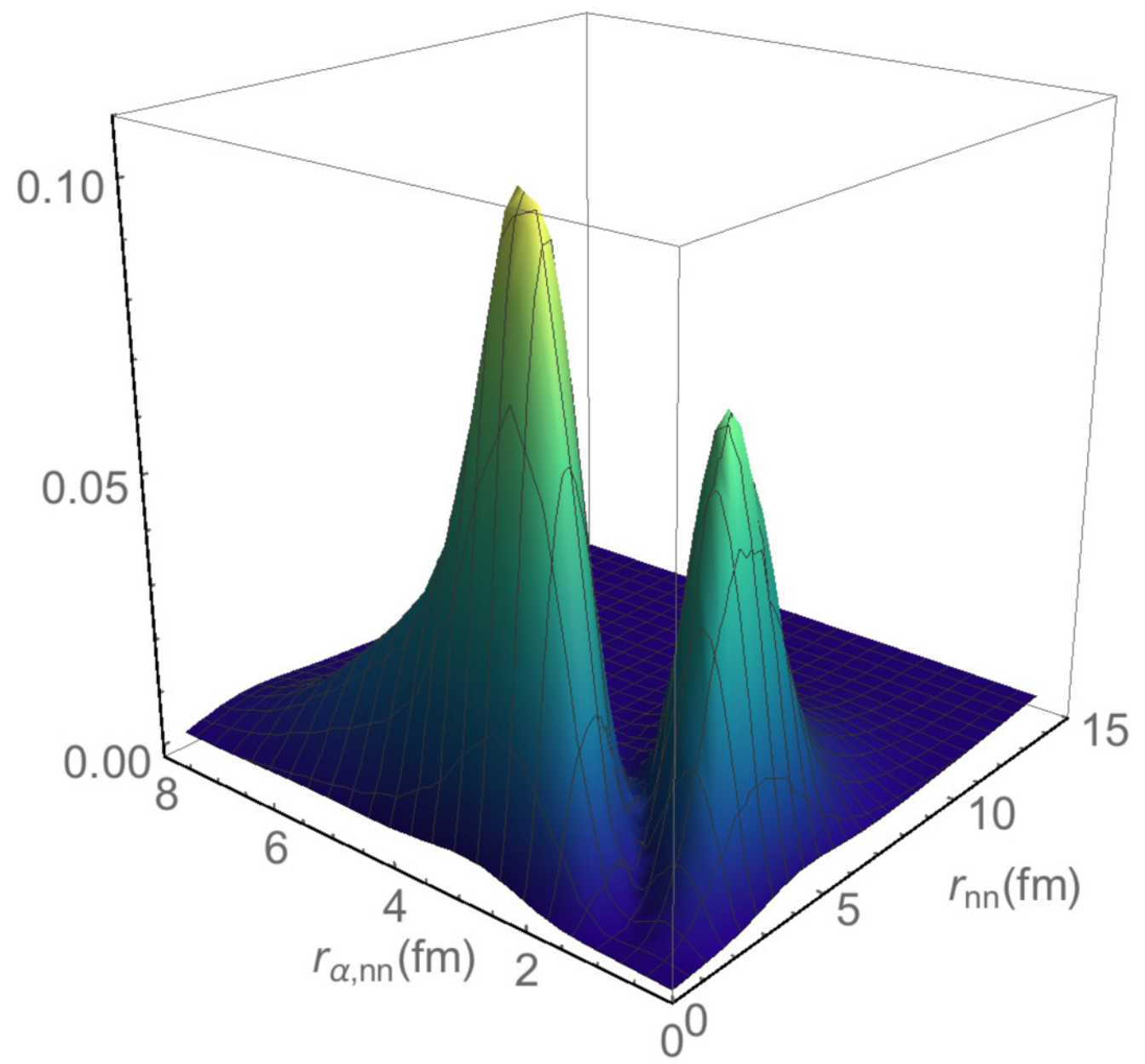

*Figure 6 Probability distribution the $J^\pi = 0^+$ ground state of the $^6$He. The $r_{nn}$ and $r_{\alpha,nn}$ are, respectively, the distance between the two neutrons and the distance between the c.m. of $^4$He and that of the two neutrons. For further details see Ref. [15].*

## Transfer reactions

*Ab initio* approaches to nuclear dynamics hold the promise to provide a more profound understanding not only of nuclear scuttering but also complex reactions. The most advanced application of the NCSMC so far is the calculation of polarized deuterium-tritium (DT) thermonuclear fusion[16]. The DT fusion, i.e. the $(d,n)$ transfer process $^3$H$(d,n)^4$He, is the most promising of the reactions that could power thermonuclear reactors of the future. This reaction, used at facilities such as ITER and NIF in the pursuit of sustained fusion energy production, is characterized by a pronounced $3/2^+$ resonance just above the DT threshold. It may lead to even more efficient energy generation if obtained in a polarized state, that is with the spins of the deuteron ($1^+$) and $^3$H ($1/2^+$) aligned. While the unpolarized DT fusion has been investigated in many experiments, very few measurements with polarized $d$ and/or $^3$H nuclei have been performed due to experimental challenges. NCSMC calculations with modern chiral NN and 3N interactions as the only input were able to demonstrate the small contributions of partial waves with orbital momentum $\ell > 0$ in the vicinity of the $3/2^+$ resonance. They predict the DT reaction rate for realistically polarized reactants and show that the reaction rate increases compared to the unpolarized one and, further, the same reaction rate as the unpolarized one can be achieved at a lower temperature (see Fig. 7).

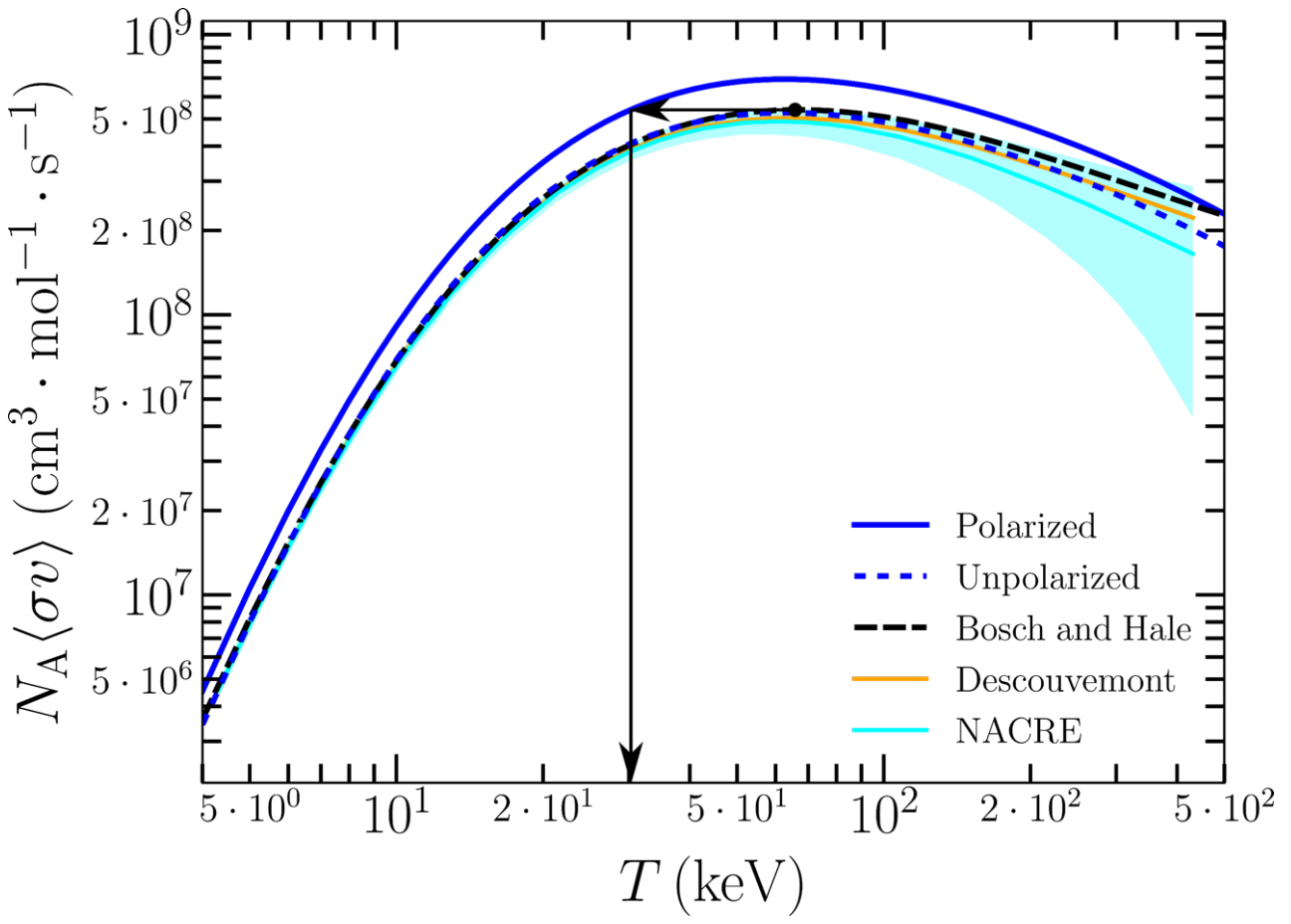

*Figure 7 NCSMC calculated $^3$H(d, n)$^4$He reaction rate with and without polarization. A realistic 80% polarization of D and T with their spins aligned is considered. The arrows in the figure show that, with polarization, a reaction rate of equivalent magnitude as the apex of the unpolarized reaction rate is reached at lower temperatures. Shown evaluations of experimental data are for unpolarized nuclei. For further details see Ref. [16].*

## The future

New technical developments of the NCSMC approach currently under way will allow calculations of reactions induced by three-nucleon projectiles, two-nucleon transfer processes, the description of $\alpha$-clustering, $\alpha$ scattering and capture processes, and $(\alpha, N)$ transfer reactions. Among our long-term goals are studies of systems with three-body clusters, e.g., Borromean two-neutron halo nuclei such as $^{11}$Li, and in general reactions with three-body final states.

Overall, *ab initio* calculations of nuclear structure and reactions made tremendous progress in the past decade and have a bright future. These calculations became feasible beyond the lightest nuclei; they make connections between low-energy quantum chromo-dynamics and many-nucleon systems and find applicability to nuclear astrophysics, nuclear reactions relevant for energy generation, as well as to evaluations of nuclear matrix elements needed in tests of fundamental symmetries and physics beyond the standard model. In synergy with experiments, *ab initio* nuclear theory is the right approach to understand low-energy properties of atomic nuclei.

## Acknowledgments

This work was supported by the NSERC Grant No. SAPIN-2016-00033 and by the U.S. Department of Energy, Office of Science, Office of Nuclear Physics, under Work Proposals No. SCW0498. TRIUMF receives federal funding via a contribution agreement with the National Research Council of Canada. This work was prepared in part by LLNL under Contract No. DE-AC52-07NA27344. Computing support came from an INCITE Award on the Titan supercomputer of the Oak Ridge Leadership Computing Facility (OLCF) at ORNL, from Westgrid and Compute Canada, and from the LLNL institutional Computing Grand Challenge Program.

[1] T. D. Morris, J. Simonis, S. R. Stroberg, C. Stumpf, G. Hagen, J. D. Holt, G. R. Jansen, T. Papenbrock, R. Roth, and A. Schwenk, Phys. Rev. Lett. **120**, 152503 (2018).

[2] K.M. Nollett, S.C. Pieper, R.B. Wiringa, J. Carlson, G.M. Hale, Phys. Rev. Lett. 99, 022502 (2007).

[3] P. Navrátil, S. Quaglioni, G. Hupin, C. Romero-Redondo, A. Calci, Physica Scripta **91**, 053002 (2016).

[4] S. Elhatisari, D. Lee, G. Rupak, E. Epelbaum, H. Krebs, T.A. L?ahde, T. Luu, U.G. Meiner, Nature **528**, 111 (2015).

[5] J. Rotureau, P. Danielewicz, G. Hagen, G. R. Jansen, and F. M. Nunes, Phys. Rev. C **98** 044625 (2018).

[6] S. Baroni, P. Navrátil, and S. Quaglioni, Phys. Rev. Lett. **110**, 022505 (2013); Phys. Rev. C **87**, 034326 (2013).

[7] B. R. Barrett, P. Navrátil, and J. P. Vary, Prog. Part. Nucl. Phys. **69**, 131 (2013).

[8] Y. C. Tang, M. LeMere and D. R. Thompsom, Phys. Rep. **47**, 167 (1978).

[9] Rimantas Lazauskas, Phys. Rev. C **97**, 044002 (2018).

[10] G. Hupin, S. Quaglioni, P. Navrátil, Phys. Rev. C **90**, 061601(R) (2014).

[11] G. Hupin, S. Quaglioni, P. Navrátil, Phys. Rev. Lett. **114**, 212502 (2015).

[12] J. Dohet-Eraly, P. Navrátil, S. Quaglioni, W. Horiuchi, G. Hupin, F. Raimondi, Phys. Lett. B **757**, 430 (2016).

[13] A. Calci, P. Navrátil, R. Roth, J. Dohet-Eraly, S. Quaglioni, G. Hupin, Phys. Rev. Lett. **117**, 242501 (2016).

[14] A. Kumar, R. Kanungo, A. Calci, P. Navrátil, A. Sanetullaev, M. Alcorta, V. Bildstein, G. Christian, B. Davids, J. Dohet-Eraly, J. Fallis, A. T. Gallant, G. Hackman, B. Hadinia, G. Hupin, S. Ishimoto, R. Krücken, A. T. Laffoley, J. Lighthall, D. Miller, S. Quaglioni, J. S. Randhawa, E. T. Rand, A. Rojas, R. Roth, A. Shotter, J. Tanaka, I. Tanihata, and C. Unsworth, "Nuclear Force Imprints Revealed on the Elastic Scattering of Protons with $^{10}$C," Phys. Rev. Lett. **118**, 262502 (2017).

[15] S. Quaglioni, C. Romero-Redondo, P. Navrátil, and G. Hupin, Phys.Rev. C **97**, 034332 (2018).

[16] G. Hupin, S. Quaglioni, and P. Navrátil, Nature Communications (2019) 10:351; https://doi.org/10.1038/s41467-018-08052-6